\documentclass[sigconf]{acmart}

\AtBeginDocument{%
  \providecommand\BibTeX{{%
    \normalfont B\kern-0.5em{\scshape i\kern-0.25em b}\kern-0.8em\TeX}}}

\setcopyright{acmcopyright}
\copyrightyear{2020}
\acmYear{2020}
\acmDOI{10.1145/1122445.1122456}

\acmConference[AVI2CH '20]{AVI2CH '20: Workshop on Advanced Visual Interfaces and Interactions in Cultural Heritage}{Sept 28 - October 2, 2020}{Ischia, Italy}
\acmBooktitle{AVI2CH '20: Workshop on Advanced Visual Interfaces and Interactions in Cultural Heritage, Ischia, Italy, Sept 28 - October 2, 2020}
\acmPrice{00.00}
\acmISBN{123}

\begin{document}

\title{From pixels to notes: a computational implementation of synaesthesia for cultural artefacts}

\author{Dimitrios Kritikos, Kostas Karpouzis}
\email{kkarpou@cs.ntua.gr}
\orcid{1234-5678-9012}
\affiliation{%
  \institution{Artificial Intelligence and Learning Systems Lab,\\National Technical University of Athens}
  \city{Athens, Greece}
}

\renewcommand{\shortauthors}{Kritikos and Karpouzis}

\begin{abstract}
Synaesthesia is a condition that enables people to sense information in the form of several senses at once. This work describes a Python implementation of a simulation of synaesthesia between listening to music and viewing a painting. Based on Scriabin's definition, we developed a deterministic process to produce a melody after processing a painting, mimicking the production of notes from colours in the field of view of persons experiencing synaesthesia.
\end{abstract}



\maketitle

\section{Introduction}
Synaesthesia \cite{Ramachandran} is a neurological condition that enables the brain to process and sense information in the form of several senses at once, despite experiencing only one or some of them. For example, a person with synaesthesia may hear sounds while also visualizing them as colours. Synaesthesia may be encountered in many forms, the most usual being \emph{chromaesthesia} \cite{Otoole2019}, where a person interprets a music or sound signal as colours. Other forms include:
\begin{itemize}
    \item Lexical-gustatory \cite{Ward2003}, where hearing words is accompanied by the sense of certain tastes
    \item Mirror-touch \cite{Ward2015}, where people may sense being touched merely being watching other people touching parts of their body
    \item Grapheme-color \cite{Jaencke2009}, where people associate letters and numbers with colours, with each number corresponding to a different colour for different persons
    \item Number-form, where people associate numbers with specific shapes or arrangements in 3D space, and
    \item Personification \cite{Smilek2007}, where people perceive letters, numbers, days or anything that can be arranged as a sequence as entity with its own personality
\end{itemize}

In general, synaesthesia may also be correlated with emotions, which makes sense since emotions (\cite{Cowie2011}, \cite{Karpouzis2016}) are triggered in specific areas of the brain \cite{Caridakis2010}, which may also participate in processes related to synaesthesia. According to Simner \cite{Simner2013}, synaesthetes comprise about 4\% of the general population; that percentage seems to be higher among artists of any kind. Synaesthesia is also probably hereditary \cite{Cytowic2011} with more than 40\% of synaesthetes having a first-degree relative experiencing the same condition. Emotion-aware mapping (\cite{Lee2016}) or matching of music  to images usually takes place  at a higher level than pixels, i.e. by incorporating shapes, places or events (\cite{Gao2018}, \cite{Chen2008}) included in an image or music genres in the process. 

This work aims to simulate the form of synaesthesia which happens when individuals are listening to music and viewing a painting. This combination is interesting, since both media are essentially based on receiving and experiencing waves, in terms of perception (\cite{Sergio2015}): sound waves collected by the human ear and light waves falling on cones and rods in the human eye and ultimately being translated to colours \cite{Waterworth1997}. There have been a number of computational approaches regarding the transformation of an image to an audio file, most notably PixelSynth \cite{Pixelsynth} and Spectrogram audio player (SAP) \cite{Spectrogram}. Pixelsynth was created by artist and coder Olivia Jack and works with monochrome images by mapping luminosity to note; its web-based implementation offers a limited selection of interactive tools (e.g. image rotation) for users to change the result of the conversion. SAP  utilises the concept of a spectrogram, i.e. a visual representation of the spectrum of frequencies of sound or other signal as they vary with time, and allows users to change the length of the output audio file or sampling density of the image.

Both of the above-mentioned implementations offer an insight to the world of synaesthesia, but the mapping they utilise is either arbitrary or based on the spectral (and not visual) representation of the image, leading to different properties of the colours being used. Based on theoretical approaches of synaesthesia, as well as mappings described by a prominent synaesthete, we developed a deterministic process to produce a melody when processing a painting, mimicking the production of notes from colours in the field of view of a person experiencing synaesthesia. Our purpose was to investigate whether different painting styles and object arrangements in paintings and drawings would produce similarly arranged music melodies. Our implementation was coded in Python, and typically takes around 5 seconds for a 1920x1080 image on an i7 laptop computer.

\section{Algorithm design}
Designing and implementing a system which mimics synaesthesia between images and music starts with identifying a correlation between each color and a note. Since we cannot fully comprehend the process which takes place in a synaesthete's brain, our best bet is to formalize their accounts of how \emph{they} perceive the connection between colors and notes.

\begin{figure}[ht]
\caption{Scriabin's mapping of notes to colors}
\includegraphics[width=8cm]{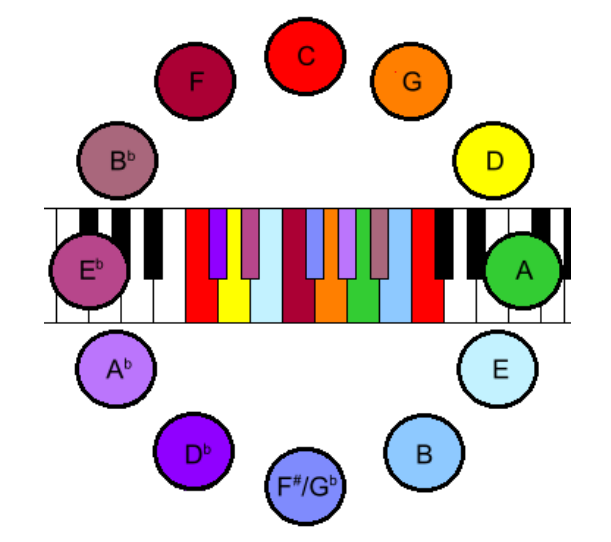}
\label{fig:scriabin}
\end{figure}

A well-known mapping of image colours to music notes is based on the work of Alexander Scriabin, a Russian composer who also authored an index of colour representations between music notes, based on his own perception of synaesthesia. In this work, we will build upon this mapping to transform a painting into a sequence of music notes, incorporating higher-level concepts, such as colour and music harmony, in the process. Besides his work in music, Scriabin produced an index which maps notes to colors, according to his own perception. This elementary mapping (see Figure \ref{fig:scriabin}) is based on elementary, saturated color tones and needs to be extended in order to capture the variations found in paintings or photographs.

A more flexible color representation utilizes 3 dimensions on the form of a color cone, where the two dimensions of the base describe the color hue (the actual color information, usually depicted with a label, e.g. `green' or `red') and the saturation (how clear a color is or, conversely, how much white or black has been used to `wash it out') and the third dimension of the cone describing the amount of light used (how bright the color is). This representation, which is quite close to how humans perceive and verbally describe color, can then be mapped to the usual RGB color encoding, found in digital representations of paintings (files or computer screens). In addition to this, and to further enhance the variety of the notes produced when processing a digital image, we also take into account color harmonies \cite{Garau1993}. There are 6 main classes of color harmonies:
\begin{itemize}
    \item Complementary colors, i.e. colors at opposite sides of the color circle (or the base of the color cone)
    \item Analogous colors, i.e. neighboring colors which differ by 30 degrees on each side (left and right on the color circle)
    \item Split-complementary colors, a triad consisting of two analogous colors and the complementary of one of them
    \item Triad colors, which form an isosceles triangle (differ by 60 degrees)
    \item Tetradic colors, formed by two pairs of complementary colors, and
    \item Square, where colors differ by 90 degrees, forming a square on the color circle or the base of the color cone
\end{itemize}
In our implementation, color harmonies are recognized by processing the image and then mapped to music chords to provide a richer tune. We also take into account the mean luminosity of each image segment: if the segment is darker than the mean luminosity of the image, we utilise a minor chord, while for a brighter segment, the chord used in in major. This is in line with our general perception of melancholic music being correlated with darker colors.

\subsection{Music tempo and volume}
Beginning by segmenting the image from left to right, the tempo of the sequence is calculated upon the saturation of colours in each segment: a slow tempo (around 75 beats per minute – bpm) is the result of low saturation colours, while higher saturation may produce tempos up to 160 bpm. 

Following from that, our algorithm chooses which notes to play for the particular image segment, taking into account the percentage of pixels in the segment which correspond to each of the primary colours, as per Scriabin’s definition: if one of those colours is found in more than 5\% of the segment pixels, then the corresponding note is included in the music sequence. In the same framework, the volume for each note is calculated with respect to the mean luminosity of the relevant pixels, while its value (i.e. duration) is based on the variety of the colours in the segment: the richer in colours, the shorter each note in the music sequence.

\section{Implementation}
\begin{figure}[ht]
\caption{Results from analyzing a complete image}
\includegraphics[width=8cm]{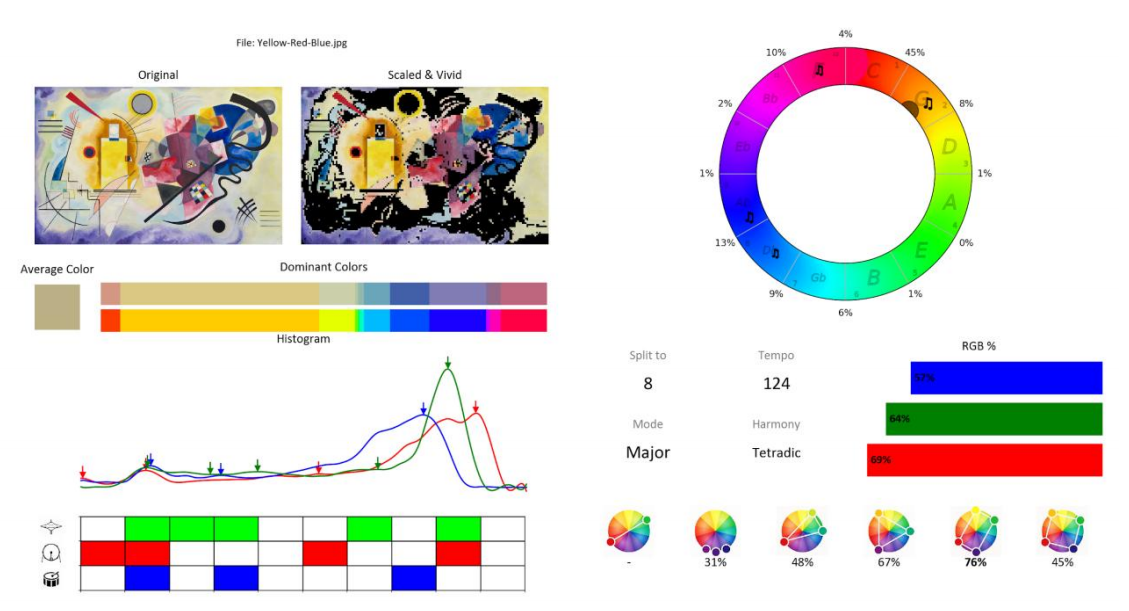}
\label{fig:image_analysis}
\end{figure}

\begin{figure}[ht]
\caption{Results from analyzing an image segment}
\includegraphics[width=8cm]{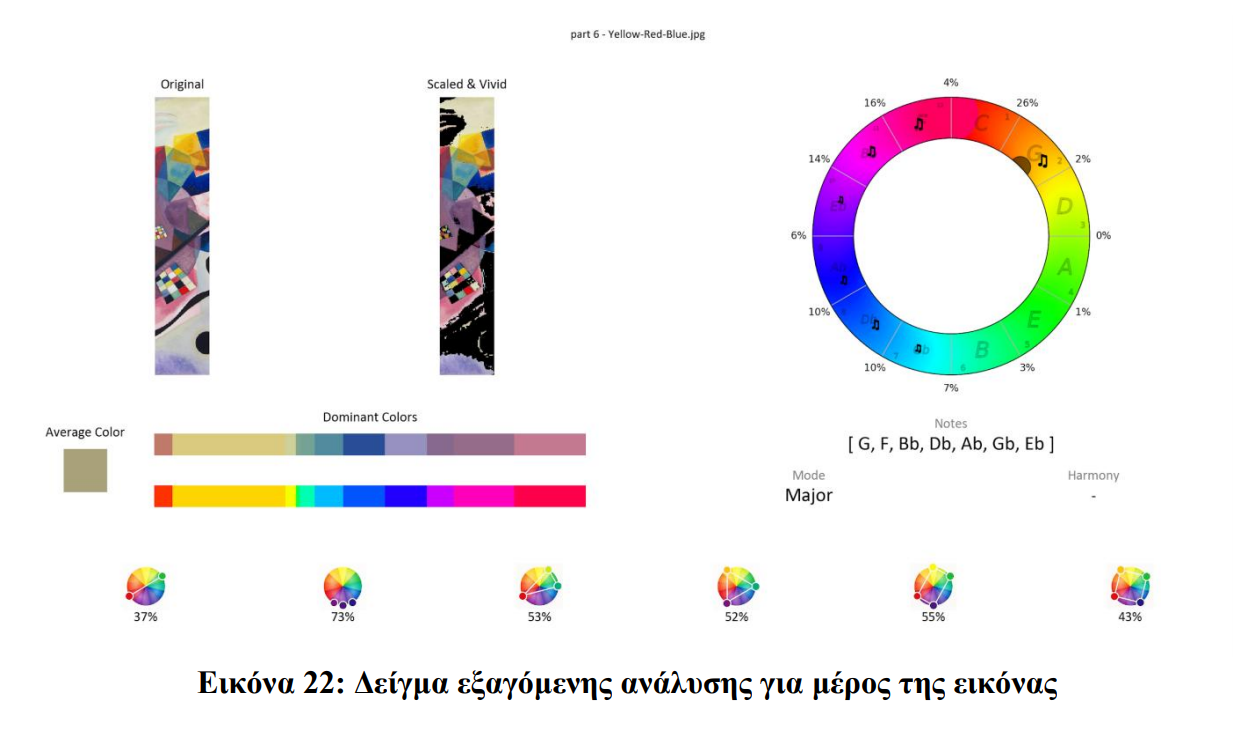}
\label{fig:segment_analysis}
\end{figure}

Figure \ref{fig:image_analysis} shows the results from analyzing a complete image, while \ref{fig:segment_analysis} shows how individual parts (segments) of an image can be processed. 

The overall pipeline is summarized in the following sequence:

\begin{enumerate}
    \item Our implementation starts with downsizing the input image (painting or photo); this step has limited effect on the actual color harmonies and richness, but results in a much faster implementation
    \item Then, the image is filtered to eliminate very bright or very dark pixels, which do not contribute to the mapping process, since they carry very little color information
    \item The remaining pixels are mapped to each of the 12 segments of the color model; the algorithm only takes into account parts of the color model with more than 5\% of the image pixels
    \item Participating colors are then ranked according to number of pixels in the image and checked for color harmonies
    \item The color information is then mapped to notes, each with a different volume, depending on the average saturation of the respective pixels: the clearer a color is, the higher the volume of the respective note
    \item Chords are calculated and possibly transformed into major or minor, according to the luminosity of the segment
    \item Finally, the melody is composed into a MIDI sequence
\end{enumerate}

This process yields a calm melody when image colors are softer and an uptempo (faster) melody for images with greater color variety. As mentioned before, darker colors result to minor chords, which matches the general conception of an imposing melody. 

\section{Conclusion}
This paper describes the implementation of a system which produces a music melody based on the colors (values and harmonies) of a painting or photo, simulating synaesthesia. This algorithm can be used to design audiovisual cultural experiences or interactive applications \cite{Malatesta2009}. The mapping process is based on input from a prominent synaesthete, who was also a composer; despite the inherent subjectiveness of the color-to-note mapping, the algorithm results in melodies which conform to our general conception of how different color values correspond to faster vs. slower melodies or richer vs simpler chords.

\bibliographystyle{ACM-Reference-Format}

\begin{thebibliography}{00}
\bibitem{Bloom} W. Bloom, D.W. Fawcett. 1975. A Textbook of Histology, 10th ed, pp. 392-410. Philadelphia: Saunders.
\bibitem{Newton1704} I. Newton (1704). Opticks: or, A Treatise of the Reflexions, Refractions, Inflexions and Colours of Light. London: the Royal Society.
\bibitem{Briggs} D. Briggs (2013). PART 7. The Dimension of Hue. Retrieved from The Dimensions of Color: http://www.huevaluechroma.com/071.php
\bibitem{Simner} J. Simner, E.M. Hubbard (2013). A brief history of synesthesia research.
\bibitem{BaronCohen} S. Baron-Cohen, L. Burt, F. Smith-Laittan, J. Harrison, P. Bolton (1996) Synaesthesia: Prevalence and familiarity. Perception, 25(9), 1073–1080. https://doi.org/10.1068/p251073
\bibitem{Barnett} K. J. Barnett, C. Finucane, J. Asher, G. Bargary, A. P. Corvin (2008) Familial patterns and the origins of individual differences in synaesthesia, Cognition. 2008 Feb;106(2):871-93
\bibitem{Ramachandran} V. S Ramachandran, E. M. Hubbard (2001). Synaesthesia: a window into perception, thought and language. Journal of consciousness studies, 8(12), 3-34.
\bibitem{Otoole2019} P. O'Toole, D. Glowinski, M. Mancini (2019). Understanding Chromaesthesia by Strengthening Auditory-Visual-Emotional Associations. In 2019 8th International Conference on Affective Computing and Intelligent Interaction (ACII) (pp. 1-7). IEEE.
\bibitem{Ward2003} J. Ward, J. Simner (2003). Lexical-gustatory synaesthesia: linguistic and conceptual factors. Cognition, 89(3), 237-261.
\bibitem{Ward2015} J. Ward and M. J. Banissy (2015) Explaining mirror-touch synesthesia. Cognitive Neuroscience 6.2-3 (2015): 118-133.
\bibitem{Jaencke2009} L. Jaencke, G. Beeli, C. Eulig, J. Haenggi (2009). The neuroanatomy of grapheme–color synesthesia. European Journal of Neuroscience, 29(6), 1287-1293.
\bibitem{Smilek2007} D. Smilek, K. A. Malcolmson, J. Carriere, M. Eller, D. Kwan, M. Reynolds (2007) When “3” is a Jerk and “E” is a king: personifying inanimate objects in synesthesia." Journal of Cognitive Neuroscience 19, no. 6 (2007): 981-992.
\bibitem{Simner2013} J. Simner and Edward M. Hubbard, eds. Oxford handbook of synesthesia. Oxford University Press, 2013.
\bibitem{Cytowic2011} R. E. Cytowic  and David M. Eagleman. Wednesday is indigo blue: Discovering the brain of synesthesia. MIT Press, 2011.
\bibitem{Garau1993} A. Garau (1993). Color harmonies. University of Chicago Press.
\bibitem{Cowie2011} R. Cowie, C. Cox, JC Martin, A. Batliner, D. Heylen, K. Karpouzis (2011) Issues in Data Labelling. In: Cowie R., Pelachaud C., Petta P. (eds) Emotion-Oriented Systems. Cognitive Technologies. Springer, Berlin, Heidelberg.
\bibitem{Karpouzis2016} K. Karpouzis and G. N. Yannakakis. Emotion in Games. Springer, 2016.
\bibitem{Malatesta2009} L. Malatesta, A. Raouzaiou, L. Pearce, K. Karpouzis. "Affective Interface Adaptations in the Musickiosk Interactive Entertainment Application." In International Conference on Intelligent Technologies for Interactive Entertainment, pp. 102-109. Springer, Berlin, Heidelberg, 2009.
\bibitem{Karpouzis2017} K. Karpouzis, Affective and Gameful Interfaces. In Proceedings of the 22nd International Conference on Intelligent User Interfaces Companion (IUI ’17 Companion). Association for Computing Machinery, New York, NY, USA, 33–34, 2017.
\bibitem{Waterworth1997} J. A. Waterworth (1997). Creativity and sensation: The case for synaesthetic media. Leonardo, 30(4), 327-330.
\bibitem{Caridakis2010} G. Caridakis, K. Karpouzis, M. Wallace, L. Kessous, N. Amir (2010). Multimodal user’s affective state analysis in naturalistic interaction. Journal on Multimodal User Interfaces, 3(1-2), 49-66.
\bibitem{Pixelsynth} Ojack.xyz., Pixelsynth [online] Available at: <https://ojack.xyz/PIXELSYNTH/> [Accessed 9 July 2020].
\bibitem{Spectrogram} Spectrogram [online] Available at: <https://nsspot.herokuapp.com/imagetoaudio/> [Accessed 9 July 2020].
\bibitem{Lee2016} T. Lee, H. Lim, D. Kim, S. Hwang, K. Yoon (2016). System for matching paintings with music based on emotions. In SIGGRAPH ASIA 2016 Technical Briefs (pp. 1-4).
\bibitem{Gao2018} Y. Gao, X. Lingyun (2018) "Aesthetics-Emotion Mapping Analysis of Music and Painting." In 2018 First Asian Conference on Affective Computing and Intelligent Interaction (ACII Asia), pp. 1-6. IEEE.
\bibitem{Chen2008} C.-H. Chen, W. Ming-Fang, J. Shyh-Kang, C. Yung-Yu (2008) "Emotion-based music visualization using photos." In International Conference on Multimedia Modeling, pp. 358-368. Springer, Berlin, Heidelberg.
\bibitem{Sergio2015} G. C. Sergio, R. Mallipeddi, J. S. Kang, M. Lee (2015) Generating music from an image. In Proceedings of the 3rd International Conference on Human-Agent Interaction (pp. 213-216).




\end{thebibliography}

\end{document}